\documentclass[journal, a4paper]{IEEEtran}
\IEEEoverridecommandlockouts

\usepackage[font=footnotesize,caption=false]{subfig} 

\usepackage{cite}
\usepackage{amsmath,amssymb,amsfonts}
\usepackage{graphicx}
\usepackage{textcomp}
\usepackage{xcolor}
\usepackage{booktabs}
\usepackage{glossaries}
\usepackage{subfig}
\usepackage{algorithm}
\usepackage{algpseudocode}
\usepackage{bbm}
\usepackage{bm}
\usepackage{multirow}
\usepackage{tabularx}
\usepackage{svg}
\usepackage[font=footnotesize, skip=2pt]{caption}

\usepackage{tikz}
\usetikzlibrary{calc}
\usetikzlibrary{shapes.arrows}
\usetikzlibrary{plotmarks}
\usetikzlibrary{patterns}
\usetikzlibrary{arrows.meta}
\usetikzlibrary{positioning}
\usetikzlibrary{spy}
\usetikzlibrary{automata}

\usepackage{pgfplots}
\usepackage{pgfplotstable}
\usepgfplotslibrary{colorbrewer}
\usepgfplotslibrary{fillbetween}

\newacronym{fm}{FM}{Foundation Model}
\newacronym{ml}{ML}{Machine Learning}
\newacronym{genai}{GenAI}{Generative AI}
\newacronym{ai}{AI}{Artificial Intelligence}
\newacronym{llm}{LLM}{Large Language Model}
\newacronym{dnn}{DNN}{Deep Neural Network}
\newacronym{aigc}{AIGC}{\gls{ai} Generated Content}
\newacronym{bpp}{BPP}{Bits Per Pixel}
\newacronym{mse}{MSE}{Mean Square Error}
\newacronym{psnr}{pSNR}{Peak Signal to Noise Ration}
\newacronym{fid}{FID}{Fréchet Inception Distance}
\newacronym{lpips}{LPIPS}{Learned Perceptual Image Patch Similarity}
\newacronym{ps}{PS}{Pixel Swapping}
\newacronym{pe}{PE}{Prompt Extension}
\newacronym{iot}{IoT}{Internet of Things}
\newacronym{mcp}{MCP}{Model Context Protocol}
\newacronym{a2a}{A2A}{Agent2Agent}

\DeclareMathOperator*{\maximize}{maximize}
\DeclareMathOperator*{\minimize}{minimize}

\def\BibTeX{{\rm B\kern-.05em{\sc i\kern-.025em b}\kern-.08em
    T\kern-.1667em\lower.7ex\hbox{E}\kern-.125emX}}
\begin{document}

\title{Initialization and Rate-Quality Functions for Generative Network Layer Protocols} 
\author{Mathias~Thorsager,~\IEEEmembership{Student Member,~IEEE,} Israel~Leyva-Mayorga,~\IEEEmembership{Member,~IEEE,} Petar~Popovski,~\IEEEmembership{Fellow,~IEEE}
\IEEEauthorblockA{\thanks{This work was supported in part by the Velux Foundation, Denmark, through the Villum Investigator Grant ``WATER'' no. 37793. Petar Popovski was supported in part by the 6G-GOALS Project under the HORIZON program no. 101139232.\\ Mathias Thorsager, Israel Leyva-Mayorga, and Petar Popovski are with the Connectivity Section, Department of Electronic Systems, Aalborg University, Aalborg Øst 9220, Denmark. (email: mdth@es.aau.dk; ilm@es.aau.dk, petarp@es.aau.dk)}}
}
\maketitle

\begin{abstract}
Generative AI (GenAI) creates full content based on compact encodings. While GenAI has been used for applications where the generated content is returned to the encoding sender, it can also extend the capacity of communication networks by transmitting compact encodings through capacity-limited links, then generating and forwarding approximations from the GenAI node to the destination. This poses the challenge of evaluating approximation quality as a function of the rate between the source and GenAI node, while accounting for the communication overhead of learning. We present a method- and modality-agnostic initialization protocol for learning rate-quality functions in GenAI-aided networks, defining three variants: source-, node-, and destination-oriented, each with different messaging flows based on where quality is measured. The protocol augments node discovery protocols (e.g., MCP, A2A) when sources lack confidence in advertised model performance.  We illustrate operation via a minimum estimation budget calculated using a distribution-free tolerance limit , and validate using a case study on image transmission under quality constraints. Results confirm the calculated budget meets the target quality requirement, with positive gains over JPEG after around 20 post-learning transmissions for a perceptual metric and more than 100 for a goal-oriented metric, providing a practical foundation for GenAI-based network compression.
\end{abstract}
\glsresetall
\section{Introduction}

The mathematical model of communication \cite{shannon1948mathematical} assumes that the data sent by the sender Alice to the destination Bob reflects accurately what is now known or not predictable by Bob. Thus, the model has an implicit assumption that Alice knows exactly what Bob does not know about her data. In terms of communication protocols, this assumption encompasses an initialization process through which Alice has discovered what exactly Bob does not know about her data. In other words, Alice knows \emph{what} and \emph{how much} (expressed as number of bits) should be sent to Bob through the communication process.

Identifying the specific data to transmit, however, gets complicated in a networked setup, where the communication between Alice and Bob is aided by intermediate nodes, referred to as routers. The assumption about the unpredictability of the data sent by Alice is extrapolated to the routers. In fact, the router's task is only to replicate the data from the incoming link to an outgoing link, not looking into the content, entropy, or predictability of the data. This modus operandi has been essential for the scalability of the internet.

Nevertheless, there are two trends that may require redefinition of the routing functionality. The \emph{first} is the definition of the communication objective way from creating a replica of the data at the source towards a more general objective of fulfilling a certain goal or attaining a certain representation at the destination. The \emph{second} is the increase in the Generative AI (GenAI) capabilities. A router with predictive and GenAI capabilities may be able to predict the data sent by the source and generate sufficiently good data to be sent to the destination. Here \emph{sufficiently good} means that the data meets certain criteria in terms of quality or achievement of a goal.

\begin{figure}[t]
\centering
\includegraphics[width=\linewidth]{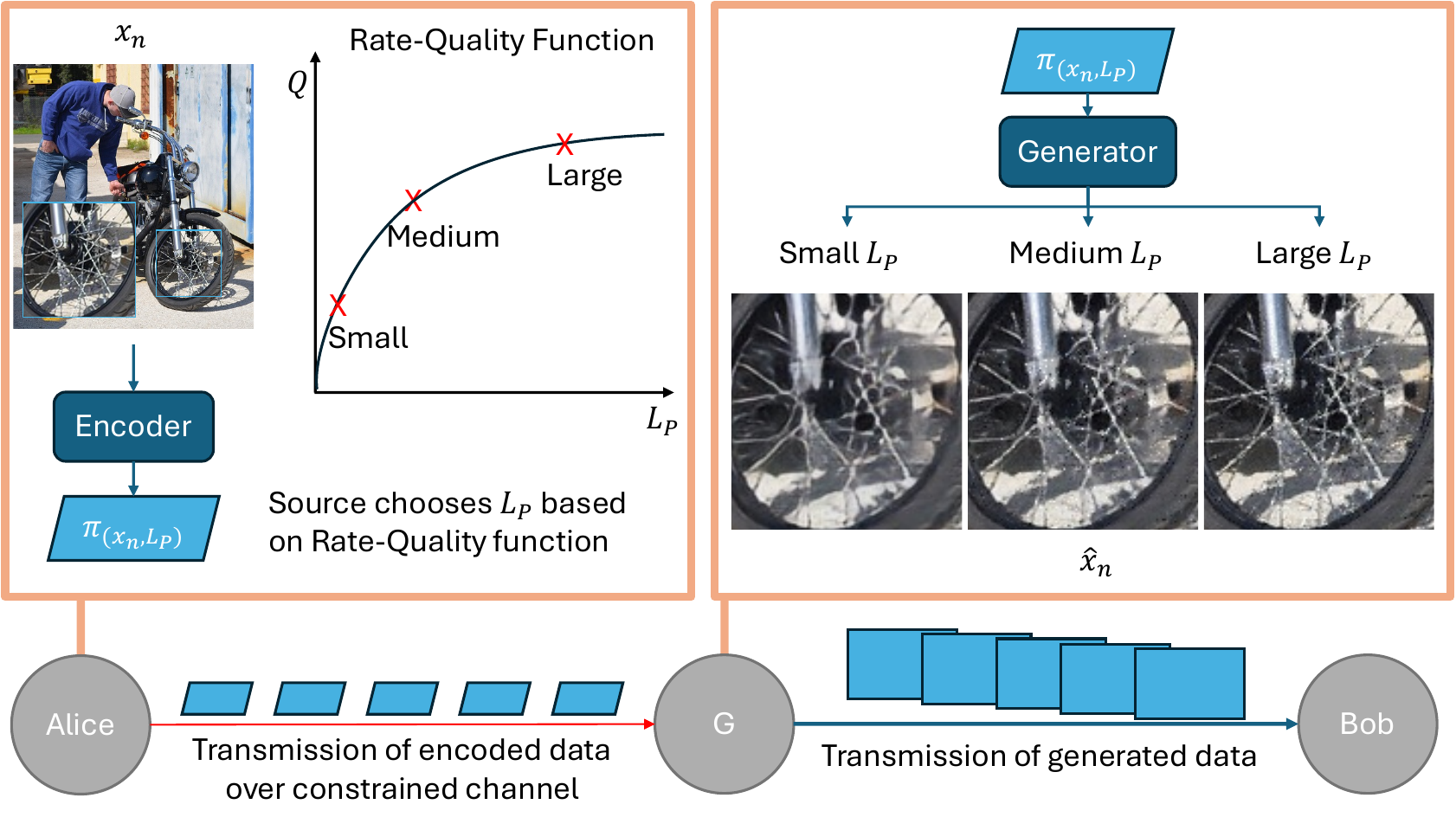}
\caption{Illustration of the communication flow using generative compression with the decoder model placed at an intermediate network node. Alice uses a local encoder to compress data and chooses the compression ratio by selecting an encoding size $L_P$ based on an estimated rate-quality function.}
\label{fig:overview} 
\end{figure}

These trends have motivated a growing body of work on integrating \gls{genai} capabilities into network infrastructure and resource management. Recognizing that existing network architectures are not designed to support the storage, computation, and transmission demands of \gls{genai} workloads, \cite{he2025advancing} propose an end-to-end programmable network framework that reallocates these resources dynamically to accommodate \gls{genai}-based services. Within such \gls{genai}-enabled networks, a central challenge is deciding how to allocate resources for individual generative requests. \cite{liu2025joint} addresses this by jointly optimizing \gls{genai} model caching and communication and computation resource allocation at the wireless edge, using a fixed, empirically fit relationship between a model's compute budget and its output quality. Similarly, \cite{mohajer2026joint} proposes a multi-agent reinforcement learning framework for SLA-aware task offloading and service orchestration under delay constraints, which could complement node selection in \gls{genai}-aided networks by prioritizing requests based on latency requirements. These works address the problem of allocating and orchestrating resources across a network of \gls{genai}-capable nodes, which is complementary to the problem addressed in this paper. Once a node has been selected, the source must characterize and exploit the specific rate-quality trade-off that node offers for its own data, rather than relying on a fixed or externally provided characterization.

In our previous work \cite{thorsager2024generative}, we introduced a general model for a \gls{genai}-aided network layer which exploits network nodes equipped with \gls{genai}-based compression models to artificially increase the data flow through capacity-constrained networks. Instead of transmitting and replicating the source data from Alice to Bob, in generative network layer we transmit encodings of the source data and generate approximations at the intermediary nodes. The key point is that the encodings are significantly smaller in size than the source data while the generated approximations score better on perceptual quality metrics as compared to traditional compression methods (such as JPEG compression) \cite{HiFiC}. This means that if the bottleneck occurs due to the capacity $c_{s,g}$ of the Alice-GenAI node rather than the capacity $c_{g,d}$ of the GenAI-destination link, such that $c_{s,d} = \min \{c_{s,g},c_{g,d}\} = c_{s,g}$, then the generative network layer can increase the reception rate of sufficiently good data.

If the goal of the communication is simply to transmit with the largest encoding size supported by the network at a certain transmission period, Alice can remain oblivious to the performance of the model used at the \gls{genai} node. However, if, as shown in Figure \ref{fig:overview}, Alice intends to minimize the use of communication or, specifically, wireless resources while taking into account the quality of the received data when choosing an encoding size, she must be able to estimate the quality of the approximated data. This requires that Alice knows, or is able to estimate, the capabilities of the \gls{genai} model at the intermediate node. In particular, the main premise of this work is that Alice must have an accurate estimate of the rate-quality function which describes the achievable quality of the approximated data given a rate (or encoding size) for the \gls{genai} model used on the source data. While the \gls{genai} node may advertise a certain performance for its generative model, e.g., through the \gls{mcp} \cite{MCP} or the \gls{a2a} protocol \cite{A2A}, Alice may not always be confident that the advertised performance will be applicable to her specific source data. If Alice is uncertain about whether her source distribution matches the one used for the advertised performance, or if she believes that there may be differences in her specific encoding strategy, she cannot be confident that she will experience the advertised performance \cite{Rate-Distortion-Perception, thorsager2024generative}.

\begin{figure*}
    \centering
    \includegraphics[width=\linewidth]{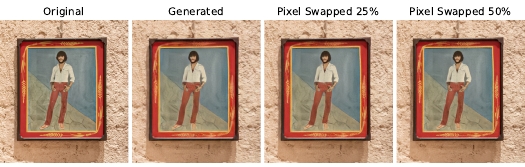}
    \caption{Example of image quality using the pixel swapping encoding method proposed in \cite{thorsager2024generative} using the HiFiC \cite{HiFiC} generative compression model. The generated image represents using the compression model alone and PixelSwapped 25\% and 50\% augments the generated image with 25 and 50 \% of pixels from the original image chosen at random. Generated, pixel swapped 25\%, and pixel swapped 50\% correspond to the small, medium, and large encoding sizes from Figure \ref{fig:overview}.}
    \label{fig:compressionShowCase}
\end{figure*}

Even with the same source data, Alice can control the encoding size through different methods, each yielding distinct rate-quality functions. In \cite{thorsager2024generative}, we introduced two such methods: \gls{pe}, which uses pre-trained encoders with varying output sizes, and \gls{ps}, which augments generated data with a fraction of the original data. The latter enables fine control over the encoding size by varying this augmentation fraction. For image generation, Alice can select a subset of pixels to transmit alongside the encoding, which node $g$ augments onto the generated image. Figures \ref{fig:overview} and \ref{fig:compressionShowCase} demonstrate this process using HiFiC \cite{HiFiC}, where 25\% and 50\% of randomly selected pixels are augmented onto the generated image, noticeably improving the visual quality in areas where significant degradation had occurred.

In addition to the differences in the source distribution and the encoding strategy, receivers may evaluate quality differently. One may consider distortion while another considers perceptual quality. Both are deviation-based metrics that do not correlate in rate-quality trade-offs \cite{Rate-Distortion-Perception}. A third receiver may evaluate quality using goal-oriented metrics based on practical downstream tasks, which may correlate poorly with either perceptual or distortion metrics \cite{SemanticMetrics}. 

Learning the rate-quality function of a \gls{genai} model is not trivial. While the optimal rate-distortion trade-off as proposed by Shannon \cite{shannon1959coding} can be solved as a closed-form expression for different data sources, not all compression methods are able to achieve this rate-quality function, especially for complex data types, such as images and video. Instead, we can measure the operational rate-distortion curve for a specific compression method, which describes the particular distortion measured for each possible rate \cite{ortega1998rate}. For traditional compression methods, e.g., JPEG, Alice would be able to perform the necessary quality measurements locally by comparing the compressed representation with the original data. \gls{genai}-based compression methods, however, require generating the approximated data from the encoding in order to measure quality, using an encoder that is specifically matched to the decoder at the \gls{genai} node. While the encoder is assumed to be small enough to be run locally by GPU-equipped UEs, the decoder generally requires substantially more computational power than neither Bob nor Alice are assumed to have. Alice must therefore rely on the network node both to generate the data needed for learning the rate-quality function and to facilitate efficient communication with Bob once that function is known. This means that the learning process involves the transmission of data to and from the network node, motivating the design of an initialization protocol.

Besides generating data for different encoding sizes, the key aspect of learning the rate-quality function is to measure the data quality. Different scenarios will warrant measuring the data at different devices. Some quality metrics may be computationally expensive and should be carried out at the node after the generation of the data. Other metrics may be goal-oriented and can only be measured by Bob based on his ability to carry out some task using the generated data. The different variants of the learning place different communication and computational loads on different parts of the network.

The main contributions of this paper are 1) a protocol and the statistical methods for defining a representative rate-quality function and its uncertainty bound and 2) a mechanism to select the optimal encoding length to achieve the desired quality with significant savings in the communication resources. 
Following a set of key assumptions and requirements on the encoder-decoder models and encoding methods, detailed in Section \ref{subsec:preconditions}, we argue that the proposed protocol is method- and modality-agnostic, providing a framework for rate-quality function estimation that enables efficient communication using various communication modes including quality-constrained, rate-constrained, and unconstrained operation. We define the required messaging flows between the source, \gls{genai} node, and destination based on which device is facilitating the learning. Furthermore, we discuss how a source can choose which GenAI node to make a contract with for data generation. We do not explore all possible learning strategies: for instance, reinforcement learning could be applied within our framework for more efficient choice of encoding sizes to transmit to the \gls{genai} node during the rate-quality estimation. Instead, we illustrate the protocol's operation through one approach based on a calculated minimum estimation budget using the distribution-free tolerance limit. We validate the protocol using two encoding approaches, two compression models, and two quality metrics, confirming that the calculated budget adheres to the target quality requirement, and show that quality-constrained operation can achieve positive communication gains after around 20 images post-learning for a perceptual quality metric and more than 100 images for an inference-based goal-oriented metric.

The remainder of the paper is structured as follows: In Section \ref{sec:System_Model}, we present the system model including the components of the rate-quality function in the context of GenAI-based compression schemes and define the communication modes which the source can use to take advantage of the \gls{genai}-based compression. In Section \ref{sec:Initialisation_protocol}, we present the initialization protocol, including how to contract a GenAI node and the different learning protocols based on which device facilitates the learning. In Section \ref{sec:InitProcedure}, we present one approach for determining the learning budget required to estimate the rate-quality function with sufficient accuracy. In Section \ref{sec:Results}, we validate the protocol's capability to estimate rate-quality functions and enable quality-constrained operation through experimental results using two distinct encoding approaches, demonstrating the trade-offs between learning cost, estimation accuracy, and communication savings. Finally, in Section \ref{sec:conclusion}, we conclude the paper.

\section{System Model}\label{sec:System_Model}

\begin{table*}[!t]
\centering
\caption{Nomenclature}
\label{tab:nomenclature}
\begin{tabular}{@{}cl@{}}
\toprule
\textbf{Symbol} & \textbf{Description} \\
\midrule
\multicolumn{2}{@{}l}{\textit{Network Elements}} \\
$s$ & Source node \\
$d$ & Destination node \\
$r$ & Traditional relay node (or set of relay nodes) \\
$g$ & GenAI-equipped node (or set of GenAI nodes) \\
$\mathcal{V}$ & Set of all nodes in the network \\
$c_{ij}$ & Capacity of path between nodes $i$ and $j$ \\
$R_{ij}$ & Maximum reliable data rate between nodes $i$ and $j$ \\
$f_{ij}$ & Traffic flow between nodes $i$ and $j$ \\
$\lambda$ & Rate of data generation at source $s$ \\
\midrule
\multicolumn{2}{@{}l}{\textit{Data and Encodings}} \\
$\mathcal{X}$ & Data set consisting of $N$ data points \\
$x_n$ & Single data point (e.g., image, video frame, audio segment), $n \in \{1,2,\dotsc,N\}$ \\
$L$ & Size of original data point \\
$N$ & Total number of data points in $\mathcal{X}$ \\
$f_\theta(\cdot)$ & Encoder model with parameters $\theta$ \\
$\pi_{(x_n,L_p)}$ & Encoding generated from data $x_n$ with size $L_p$ \\
$L_p$ & Encoding size (in bits per pixel for images) \\
$L_\text{min}$ & Minimum encoding size \\
$L_\text{avg}$ & Average encoding size over subset $\mathcal{P}_{N_p}$ \\
$\mathcal{P}_{N_p}$ & Subset of $N_p$ encoding sizes used in learning \\
$N_p$ & Number of encoding sizes used in learning, $N_p = |\mathcal{P}_{N_p}|$ \\
$g_\theta(\cdot)$ & Generative model with parameters $\theta$ \\
$\hat{x}_n$ & Generated approximation of data point $x_n$ \\
\midrule
\multicolumn{2}{@{}l}{\textit{Quality and Rate-Quality Function}} \\
$Q(\hat{x}_n)$ & Quality of generated approximation $\hat{x}_n$ \\
$D(L_p, g_\theta(\cdot))$ & Rate-quality function for generative model $g_\theta$ \\
$\delta(x_n,\hat{x}_n)$ & Deviation-based quality metric between original $x_n$ and generated $\hat{x}_n$ \\
$\hat{Q}(L_p)$ & Estimated quality as random variable for encoding size $L_p$ \\
$p$ & Target coverage of the tolerance bound \\
$\gamma$ & Confidence level that the tolerance bound achieves coverage $p$ \\
$Q_\text{min}$ & Minimum required quality constraint \\
$\alpha^*$ & Target probability for quality adherence \\
\midrule
\multicolumn{2}{@{}l}{\textit{System Latency}} \\
$T_L$&Latency for processing and transmitting a single data point, $T_L = T_P + T_C + T_G$\\
$T_P$&Latency for generating encoding at source \\
$T_C$&Latency for transmitting encoding from source to node $g$ \\
$T_G$&Latency for generating data approximation at node $g$ \\
\midrule
\multicolumn{2}{@{}l}{\textit{Learning Process}} \\
$N_L$ & Number of data points used for learning rate-quality function \\
$N_{\min}$ & Minimum estimation budget required by the tolerance bound \\
$\kappa_s$ & Communication cost per data point for source-oriented learning \\
$\kappa_n$ & Communication cost per data point for node-oriented learning \\
$\kappa_d$ & Communication cost per data point for destination-oriented learning \\
$K$ & Total learning cost, $K = N_L \kappa_i$ where $i \in \{s,n,d\}$ \\
\midrule
\multicolumn{2}{@{}l}{\textit{Operational Transmission}} \\
$N_C$ & Number of data points transmitted in post-learning phase \\
$L_P^*$ & Optimal encoding size \\
$w$ & Per image communication savings, $w=(|x_n| - L_P^*)$ \\
$W$ & Total communication savings, $W = N_Cw$ \\
\bottomrule
\end{tabular}
\end{table*}

We consider a network with a source node $s$, a destination $d$, a subnetwork of traditional relay nodes $r$, and a subnetwork of relay nodes with \gls{genai} capabilities $g$. Data is transmitted from $s$ to $d$ through either $r$ or $g$. Let $\mathcal{V}$ be the set of all nodes. Each of the edges $(i,j)$ in the network topology represents an $i-j$ path containing one or more hops between these nodes, which operates under a traditional packet routing model. Similarly, $r$ represents the nodes along the shortest path between $s$ and $d$ that are independent of the $s-g$ and $g-d$ paths. Therefore, the capacity $c_{ij}$ represents the capacity of the single- or multi-hop path between a pair of nodes. $R_{i,j}$ and $f_{i,j}$ are the maximal data rates for reliable communication and the traffic flows between node $i$ and $j$, respectively.

The source $s$ attempts to transmit data $\mathcal{X}$ consisting of $N$ data points, to the destination $d$. Each data point, indexed by $x_n$ for $n\in\{1,2,\dotsc, N\}$, represents a single piece of content (e.g., an image, video frame, or audio segment). However, instead of transmitting the original data $\mathcal{X}$ through $r$, it employs one (or more) of the \gls{genai} models at the nodes in $g$ to transmit encodings on the data with significantly lower data size than the original data. The encodings are used at $g$ to generate approximations of the original data, which are then transmitted to $d$.

Transmitting the encodings instead of the original data can significantly reduce the amount of data transmitted between $s$ and $g$. However, this reduction is not without cost, since encoding the data and generating approximations require substantial computational resources, particularly when scaled beyond single-user scenarios. The proposed framework is therefore best suited to deployment regimes in which computational resources can be traded for communication gains, arising either from constrained and costly communication resources or from an abundance of computational capacity.

Despite the increased computational requirement for encoding and decoding at the source and node $g$, the resulting computational latency is likely to be comparable to existing communication latency, so this work focuses on communication cost rather than latency or queueing, which we leave for future investigation. Using the HiFiC encoder and decoder \cite{HiFiC} as an approximate example, encoding and decoding time is on the order of tens of milliseconds. The decoder model has been measured empirically at around 30 ms on an A100 \cite{careil2024towards}, and a similar GPU utilization on a low-end laptop GPU (e.g., an RTX 3050 Mobile) yields an encoder latency of around 40 ms. Against typical 3GPP use-case latency targets of 100-300 ms for cloud-offloaded AI tasks (e.g., use cases 6.28 and 6.49 \cite{3gpp_tr22870}), these processing times do not present a significant limitation. Moreover, since these figures reflect single-image encoding and decoding, GPU utilization remains low\footnote{Considering the encoder and decoder networks and the theoretical compute limit of the GPUs, we calculate an approximate utilization of less than 5\%.}, leaving substantial headroom for batch inference when the framework is scaled to multiple users sharing the same generative nodes.

\subsection{Encoding}\label{subsec:encoding}
While there are multiple ways of designing encoding schemes, including explicit and implicit encoding\footnote{In \cite{thorsager2024generative}, we used the general term prompt to refer to the encodings of this work.} as presented in \cite{thorsager2024generative}, we focus on explicit encodings in this work.

For each data point $x_n \in \mathcal{X}$, let $\pi_{(x_n,L_p)}$ be an explicit encoding generated from a function $f_\theta(x_n,L_p)$ based on the individual data point and the desired encoding size $L_p$. Here, parameter $\theta$, which is shared by the decoder model $g_\theta(\cdot)$, indicates that the encoder and decoder models are jointly trained. We assume that the encoder and decoder models are trained for a single data modality, e.g., image data or sound data. 
For explicit encoding, $s$ transmits a single encoding $\pi_{(x_n,L_p)}$ for each data point $x_n$, which is used at $g$ to generate an approximation $\hat{x}_n$ using a generative model $g_\theta(\cdot)$, i.e., $\hat{x}_n = g_\theta(\pi_{(x_n,L_p)}) = g_\theta(f_\theta(x_n,L_p))$. The set of all approximations is denoted as $\hat{\mathcal{X}} = \{\hat{x}_n : n \in \{1,2,\dotsc,N\}\}$.

\subsection{Data Quality} \label{subsec:data_quality}

Let $Q(\hat{x}_n)$ denote the quality of the approximated data $\hat{x}_n$. While the quality of the generated data is mostly defined as an actual or perceived deviation from the original data, we differentiate goal-oriented metrics from these as they do not necessarily correlate well with deviation-based metrics. Additionally, while both types of metrics assess the quality of the generated data, their numerical interpretations are not always aligned. In goal-oriented metrics, higher values typically indicate better quality (e.g., task success rate), whereas in deviation-based metrics, lower values correspond to higher quality (e.g., distortion). As such, to unify the interpretation of quality, we adjust the definition of deviation-based metrics.

Let $\delta(x_n,\hat{x}_n)$ denote the distance between the original data and the approximated data. The distance can be based on a distortion measure $\delta_D(x_n,\hat{x}_n)$ or a perceptual quality measure $\delta_P(x_n,\hat{x}_n)$. For deviation-based metrics, the quality is then defined as $Q_{\text{dev}}(\hat{x}_n) = \frac{1}{\delta(x_n,\hat{x}_n)}$, which requires both the original and approximated data for evaluation. In contrast, for goal-oriented metrics, the quality is defined as $Q_{\text{goal}}(\hat{x}_n) = f_{\text{task}}(\hat{x}_n)$, where $f_{\text{task}}(\cdot)$ measures the effectiveness of the approximated data in achieving some goal or carrying out a specific task at $d$. This could be the probability that an inference task is successful or the probability that a correct action is taken based on the approximated data \cite{SemanticMetrics}.

Regardless of the quality measure, the output of many \gls{genai} models is inherently stochastic. This means that the same encoding $\pi_{(x_n,L_p)}$ will produce different data each time it is used with the same model. While the output of the models can be made deterministic through seeding the model, it is not possible to predict what the exact output will be given a random seed, nor is it possible to design an encoding scheme that always produces perfect replicas of the original data. Instead, the encoding scheme should be designed to have the quality be correlated with the size of the encoding.

\subsection{Rate-Quality Function}\label{subsec:rate-quality_function}
Since the encodings are designed to have increasing quality as the size of the encoding increases, we define a rate-quality function 
\begin{equation}
    D(L_p,g_\theta({\cdot})) = E_\mathcal{X}[Q(g_\theta(f_\theta(x_n,L_p))],
\end{equation}
which describes the correlation between the expected quality of the approximated data generated from an encoding of size $L_p$ given a \gls{genai} model parameterized by $\theta$ at $g$ and the specific source dataset $\mathcal{X}$. It is important to note that the dependency on the model parameters $\theta$ and the source data $\mathcal{X}$, means that a learned rate-quality function does not generalize to cases of mismatching encoder and decoders, nor for cases of source data distribution shift. However, for the later case, we refer to Section \ref{subsec:pilots} where we discuss mitigation strategies for such distribution shifts occurring after the rate-quality function has been learned. 

We define the rate-quality function using the expected quality based on the fact that the data generated from \gls{genai}-based compression methods is inherently stochastic. Furthermore, we assume that neither the source nor the \gls{genai}-aided node knows the true rate-quality function for the particular data at the source. Instead, the source has to estimate the function from observed data. In this work, we estimate the rate-quality function by fitting a regression model to observed quality measurements across different encoding sizes, with model parameters estimated by minimizing the sum of squared residuals. However, if the particular data type or quality metric does not exhibit a well-behaved relationship with encoding size, machine learning-based regression methods may be necessary to accurately capture the rate-quality function.

\subsection{Communication Modes}\label{subsec:communication_modes}

Communicating efficiently using \gls{genai} nodes depends on selecting an appropriate encoding size that balances rate and quality. To formalize this trade-off, we define three communication modes, each corresponding to a distinct scenario: (1) quality-constrained, (2) rate-constrained, and (3) unconstrained. Each mode makes a different assumption about network capacity, stated in the relevant subsections below: Quality-constrained and unconstrained communication assume sufficient network capacity throughout the network, so the framework's benefit in these modes is resource efficiency rather than resolving a capacity constraint. Rate-constrained communication instead assumes a binding capacity constraint, which the framework can resolve only if the network's min-cut is located between $s$ and $g$. In all three modes, network conditions elsewhere in the topology do not enter the optimization. For each mode, we specify an optimization problem that guides the source in selecting the optimal encoding size based on an estimated rate-quality trade-off.

\subsubsection{Quality-Constrained Communication}

The goal in quality-constrained communication is to transmit with the lowest possible encoding size that will result in minimizing the traffic load while generating data of sufficient quality. Therefore, the only constraint on the source is to ensure that the data received at the destination meets the required quality, and it is assumed that the capacity of the network is able to support the transmission of the original data. Since the rate-quality function is stochastic, the only way to guarantee that the source will meet a quality constraint with probability $1$ is by transmitting the original data. Furthermore, the source will be solving the optimization function based on an estimated rate-quality function, whose accuracy depends on the amount of data used in the estimate. As such, it follows that the chosen encoding size is not guaranteed to be optimal and that the source should take into account the uncertainty in the rate-quality function by choosing larger encoding sizes relative to the amount of uncertainty in the estimate. While the source cannot determine the exact uncertainty in any given estimate of the rate-quality function, it can utilize tools, such as the prediction interval or tolerance limit, to estimate the level of uncertainty based on the number of data points used for the estimate \cite{madsen2010introduction, wilks1941determination}. To accommodate this, we define the quality requirement as $E \left[p(Q(L_p) \geq Q_\text{min}) \right] \geq \alpha^*$, where $Q_\text{min}$ is the desired quality and $\alpha^*$ is the desired probability of achieving sufficient quality. Building on these, we formulate the following constrained optimization problem.

\begin{subequations}\label{eq:quality_constrained_Mode}
\begin{IEEEeqnarray}{Cll}
\minimize_{L_p}\quad & L_p \\
\text{subject to } 
& E \left[ p(Q(L_p) \geq Q_\text{min}) \right] \geq \alpha^*.
\end{IEEEeqnarray}
\end{subequations}

\subsubsection{Rate-Constrained Communication}

For rate-constrained communication, we consider a case where data is generated at the source $s$ at a rate $\lambda$ and where each data point is $L$ bits, such that the rate $\lambda L$ would surpass the capacity of the relay network $r$. Consequently, for rate-constrained communication, it is assumed that the flow $f_{s,d}=\lambda L$, when $g$ is not utilized, is larger than the min-cut $f'_{s,d}$. This means that the divergence of each node $y_i=\sum_{i\neq j}f_{ij}-\sum_{i \neq j}f_{ji} < 0, i \in \mathcal{V}\setminus \{s,d\}$ is less than zero, which results in an unstable network. However, when the \gls{genai}-aided node $g$ is used, we have
\begin{equation}
    y_g=\sum_{j\in\mathcal{V}\setminus g}f_{gj}-\sum_{j\in\mathcal{V}\setminus g}f_{jg} \geq 0, \quad \iff f_{sg} \geq \lambda L_\text{min}
\end{equation}
 where $L_\text{min}$ is the lowest possible encoding size given the specific encoding strategy and \gls{genai} model. That is, as long as the min-cut of the network is between the source $s$ and the node $g$ and the flow $f_{gd} \leq f'_{gd}$, we can overcome the rate-constraint of the network using \gls{genai}. 

The choice of encoding size in rate-constrained communication is found through the following optimization function:
\begin{subequations}\label{eq:rate-limited_Mode}
\begin{IEEEeqnarray}{cll}
\maximize_{L_p}\quad & y_g - y_g w \frac{1}{Q(L_p)} \\
\text{subject to } 
&f_{si} \leq c_{si} & \forall i \in \mathcal{V}\setminus s\\
&f_{id} \leq c_{id} & \forall i \in \mathcal{V}\setminus d\\
&\lambda L_p \geq \lambda L_\text{min}\\
&\lambda L = f_{gd}.
\end{IEEEeqnarray}
\end{subequations}
where $w$ is a scalar that determines the importance of the quality of the generated data.

\subsubsection{Unconstrained Communication}

Even in scenarios where the network has sufficient capacity ($f_{s,d}=\lambda L$ without use of $g$ is supported) and quality requirements are absent, the source may still utilize the rate-quality function to balance encoding size against estimated quality. In particular, by removing the rate and quality constraints, the source chooses an optimal encoding size by solving the following weighted optimization problem:

\begin{subequations}\label{eq:Unconstrained_Mode}
\begin{IEEEeqnarray}{Cll}
\maximize_{L_p}\quad & y_g - y_g w \frac{1}{Q(L_p)} \\
\text{subject to }
&L_p \geq L_\text{min}
\end{IEEEeqnarray}
\end{subequations}

\subsection{System Latency}\label{subsec:latency_model}
The learning process and subsequent post-learning communication involve latency components which affect the initialization procedure (see Section \ref{sec:Initialisation_protocol}). We model the total latency for processing and transmitting a single data point as:
\begin{equation}\label{eq:latency_model}
T_L = T_P + T_G + T_C,
\end{equation}
where $T_P$ is the time needed for encoding generation at $s$, $T_G$ is the generation time of the data approximation at $g$, and $T_C$ is the communication delay for data transmission between devices ($s$, $g$, and $d$),

The computational delay at the source $s$ consists of the time to encode each data point $x_n$ into its encoding $\pi_{(x_n,L_p)}$ using the encoder $f_\theta(\cdot)$. The generation time $T_G$ represents the computational delay at node $g$ when generating the data approximation $\hat{x}_n$ using $g_\theta(\cdot)$. 

The communication delay $T_C$ depends on the amount of data transmitted to and from the source $s$ and the data rate of the links (the transmission delay), as well as the distance between the nodes (the propagation delay). During the learning phase, the exact amount of data transmitted between the devices depends on the learning protocol used by the source (see Section \ref{subsec:Learning}).

\subsection{Preconditions} \label{subsec:preconditions}
This subsection gathers the assumptions and requirements that must hold for the proposed framework to function. In line with the framework's design goal of being agnostic to the specific data, encoding method, and generative model used, these are stated as general requirements rather than tied to any particular choice of these.

The first requirement states that the encoder and decoder must be jointly trained and share model parameters $\theta$. While model drift may not be detrimental for the framework, we leave the investigation of the impact of this to future work and assume that no such drift occurs in this work.

The second, and main requirement, states that the encoding methods must be designed such that the quality is correlated with encoding size. This is importantly weaker than requiring that the rate-quality function is monotonically increasing in quality with respect to encoding size. While a non-monotonous rate-quality function will reduce the performance of the framework, it does not prevent the source from using it. In the case of quality constrained communication, as is shown in Section \ref{subsec:CaseStudy_commsSavings}, non-monotonicity in the uncertainty bound will result in unusable regions on the rate-quality function which can limit the available encoding sizes. On the other hand, for the rate-constrained and unconstrained communication modes, non-monotonicity in the mean rate-quality function can result in non-convex optimization resulting in there being no closed-form solution guarantee. This requirement also impacts the claim that the framework is agnostic to data modality. While different data modalities will require different compression models and encoding methods with varying compression performance, as long as the encoding methods are designed according to this requirement, the framework is agnostic to data modality.

\section{Learning the Rate-Quality Function}\label{sec:Initialisation_protocol}

Before the source can utilize the generative power of the \gls{genai}-aided node, it must learn the rate-quality function related to the data that will be transmitted. Since the decoder cannot be run locally, the source needs to make use of a dedicated \gls{genai} node during the initialization process, which can be split into two parts: 1) making a contract with a \gls{genai} node and 2) learning the rate-quality function by transmitting encodings to the node. These are described in the following.\footnote{The initialization process does not differ for the communication modes defined in Section \ref{subsec:communication_modes}} For contracting specifically, we present the general considerations involved in selecting a \gls{genai} node, rather than a detailed analysis of the underlying negotiations and resource-allocation mechanisms.

\subsection{Contracting a GenAI Node}\label{subsec:conctract}

The first step of the initialization process is to make a contract with a \gls{genai} node, which specifies the details and requirements for the generation of data. In the simplest case, where we only consider a single source, a single \gls{genai} node, and a single destination, it is clear that the source must either use the \gls{genai} node or transmit the data directly to the destination. However, in the case that multiple nodes are equipped with \gls{genai} models and multiple sources are actively carrying out the initialization process or utilizing the nodes for generative content transmission, the contract should specify which node(s) will be used. Whether a single or multiple nodes should be utilized depends on the specific requirements of the source and the type of data to transmit. 

The goal of the contract is to find the node(s) that are best suited to serve the source. The source evaluates a node based on KPIs that are important to the source. The main factors that a source would consider include the generation accuracy and the generation latency. Finding the optimal node is done in two steps: 1) node discovery and 2) node probing.

\subsubsection{Node Discovery}

\begin{figure}
    \centering
    \includegraphics[width=1\linewidth]{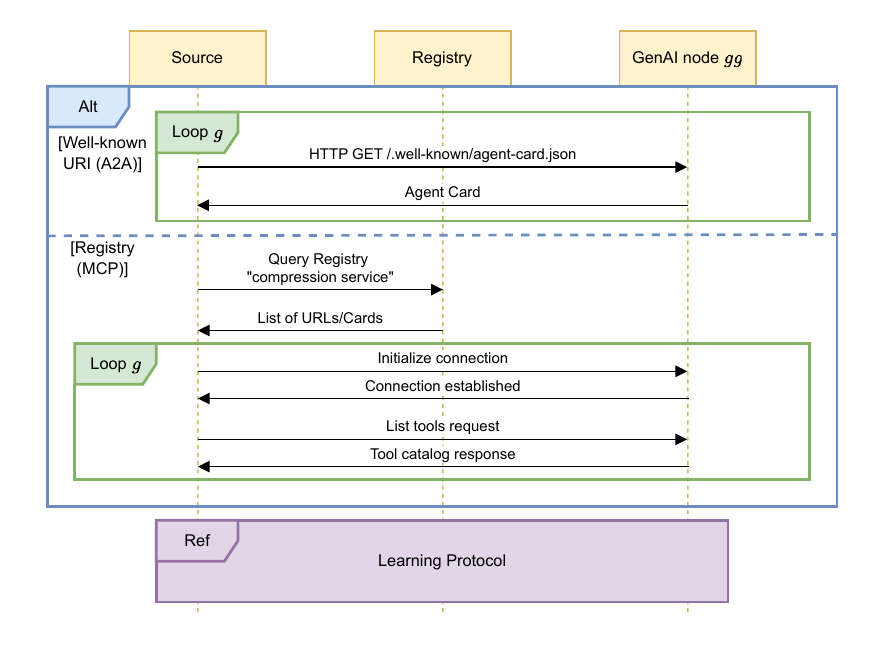}
    \caption{Message sequence diagram for the node discovery following the MCP and A2A protocols. The reference to the learning protocol indicates the beginning of one of the learning protocols presented in Section \ref{subsec:Learning} (Figures \ref{fig:InitProtc_source-oriented}, \ref{fig:InitProtc_node-oriented}, and \ref{fig:InitProtc_destination-oriented}).}
    \label{fig:InitProtc_Node_Discovery}
\end{figure}

Node discovery is a two-part process: identifying available nodes and querying the tools each provides. Through protocols such as \gls{a2a} \cite{A2A} and \gls{mcp} \cite{MCP}, the source discovers an initial set of nodes based on query options such as node location and supported data modality. The complete flow for the node discovery is shown in Fig.~\ref{fig:InitProtc_Node_Discovery}.

A2A and MCP each support different protocols for node discovery. For A2A, the standard is to use well-known URI and programmatically search for the existence of agent cards, which list the capabilities of the agent. This solves both the discovery of which nodes exist and gives the source additional information on the type of model hosted by the node. For MCP, the standard is to use a public registry of models, which will provide the URLs of MCP servers. This requires that the source additionally initialize a connection to each server in order to get information on the model capabilities.

If the source can be confident in the advertised performance of a model, then it can make the final decision on which node to make a contract with and skip the probing step. However, while a \gls{genai} model may support the specific data modality of the source, this does not guarantee that all source data will produce the same rate-quality function. The rate-quality function will be dependent on the distribution of the source data, as well as the encoding strategy used by the source. This means that the source will require additional information from the considered nodes.

\subsubsection{Probing for Node Capabilities}\label{subsubsec:NodeProbing}

In order to determine the capabilities of a node, the source must probe it and measure the relevant metrics, e.g., generation latency and generation accuracy. The probing involves transmitting encodings to the node and measuring the quality of the generated data, following the protocols described in Section \ref{subsec:Learning}. The generation latency can either be inferred from the tool or agent discovery or estimated based on an encoding (assuming that the computation time is deterministic based on the reserved resources).\footnote{While we acknowledge that the assumption of deterministic computation time does not hold in practice, accurately modeling the distribution of the computation time is beyond the scope of this paper.} However, the generation accuracy is not as simple. While a source can estimate the generation accuracy of a node based on a single encoding from a single data point, this will result in a less accurate choice of node for the contract. Instead, the source can choose to probe several times and make a better estimate of the generation accuracy. At the extreme, the source may choose to learn the entire rate-quality function of the node in the probing phase before making a contract with any node. This will result in the most accurate choice of node but will also come at a significant cost, as the source will be learning several complete rate-quality functions for different nodes. As such, there is a trade-off between the accuracy in the cost function and the cost of probing the node capabilities. However, regardless of how much probing the source does, the process of learning the rate-quality function does not change. Beyond the initial selection of a node, whether a contracted node continues to deliver as agreed over time is a separate question. This can be addressed through self-enforcement, where the source continues to probe the node periodically after contracting, coinciding with the pilot transmissions used post-learning to check for distribution shift in the source data, as described in Section \ref{subsec:pilots}, or through verifiable inference from the node, without relying on external trust. Alternatively, third-party enforcement by a network operator or other validating entity can offer broader oversight across sources and the ability to act against a node beyond a single source withdrawing its contract.

\subsection{Three Generic Protocols}\label{subsec:Learning}

Learning the rate-quality function requires transmitting encodings of varying sizes for $N_L$ data points to measure quality and estimate $D(L_p, g_\theta(\cdot))$. In particular, the rate-quality function is estimated using an arbitrary subset $\mathcal{P}_{N_p}$ of $N_p$ encoding sizes from the set of all feasible encoding sizes, where $|\mathcal{P}_{N_p}| = N_p$. 

We present three learning protocols that are differentiated by the device that measures quality and fits the rate-quality function. Table \ref{tab:learning_protocols} summarizes the key distinctions. The learned rate-quality function is used to solve the optimization problems from Section~\ref{subsec:communication_modes} during data transmission.

Each protocol can operate in two modes. In \textit{pre-transmission learning}, the source completes all $N_L$ learning iterations before transmitting operational data to $d$, ensuring all transmitted data uses the fully learned $D(L_p, g_\theta(\cdot))$. In \textit{real-time learning}, generated approximations $\hat{x}_n$ are forwarded to $d$ during the learning phase, reducing latency at the cost of transmitting data based on intermediate (less accurate) estimates of the rate-quality function. How the source can further refine the rate-quality function during the operational phase is discussed in Section \ref{subsec:pilots}.

\begin{table*}[ht]
\centering
\caption{Comparison of Learning Protocols}
\label{tab:learning_protocols}
\begin{tabular}{@{}lccc@{}}
\toprule
\textbf{Attribute} & \textbf{Source-Oriented} & \textbf{Node-Oriented} & \textbf{Destination-Oriented} \\
\midrule
Node measuring quality
& $s$ & $g$ & $d$ \\
RQ Function Fitter & $s$ & $s$ or $g$ & $s$ or $d$ \\
Original Data Tx & None & $s \to g$ & ($s \to d$) \\
Generated Data Tx & $g \to s$ & None & $g \to d$ \\
Quality Metric Compatibility & Deviation-based & Deviation-based & Goal-oriented preferred \\
Communication Cost Per Data Point & $|\hat{x}_n| + L_p$ (per loop) & $|x_n| + L_\text{min}$ (per data point) & $|\hat{x}_n| + L_p$ (per loop) \\
\bottomrule
\end{tabular}
\end{table*}

\subsubsection{Source-Oriented Learning}
\begin{figure}
    \centering
    \includegraphics[width=1\linewidth]{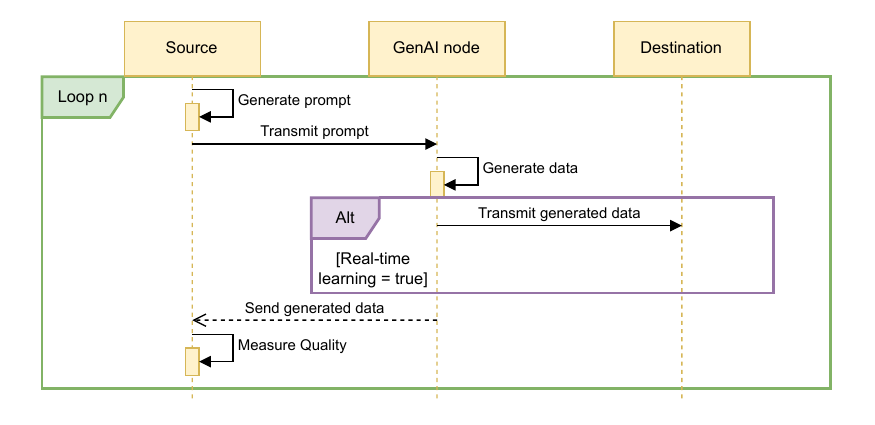}
    \caption{Message sequence diagram for the source-oriented learning process. For each of $N_L$ data points, $N_p = |\mathcal{P}_{N_p}|$ loops are required to measure the quality for the selected encoding sizes.}
    \label{fig:InitProtc_source-oriented}
\end{figure}

The source-oriented learning protocol provides the source with direct control over quality measurement and function estimation. Fig.~\ref{fig:InitProtc_source-oriented} shows the message sequence of the protocol. For each data point $x_n$, the source generates encodings $\pi_{(x_n,L_p)}$ of varying sizes $L_p \in \mathcal{P}_{N_p}$, transmits them to $g$, receives a generated approximation $\hat{x}_n$ for each encoding size, and measures quality $Q(\hat{x}_n)$ to fit $D(L_p, g_\theta(\cdot))$. This protocol is restricted to deviation-based quality metrics (Section~\ref{subsec:data_quality}), as the source cannot execute the destination's tasks to measure goal-oriented quality. The communication cost per data point is $\kappa_s = \sum_{L_p \in \mathcal{P}_{N_p}}(L_p + |\hat{x}_n|)$ for $N_p$ tested encoding sizes.

\subsubsection{Node-Oriented Learning}

\begin{figure}
    \centering
    \includegraphics[width=1\linewidth]{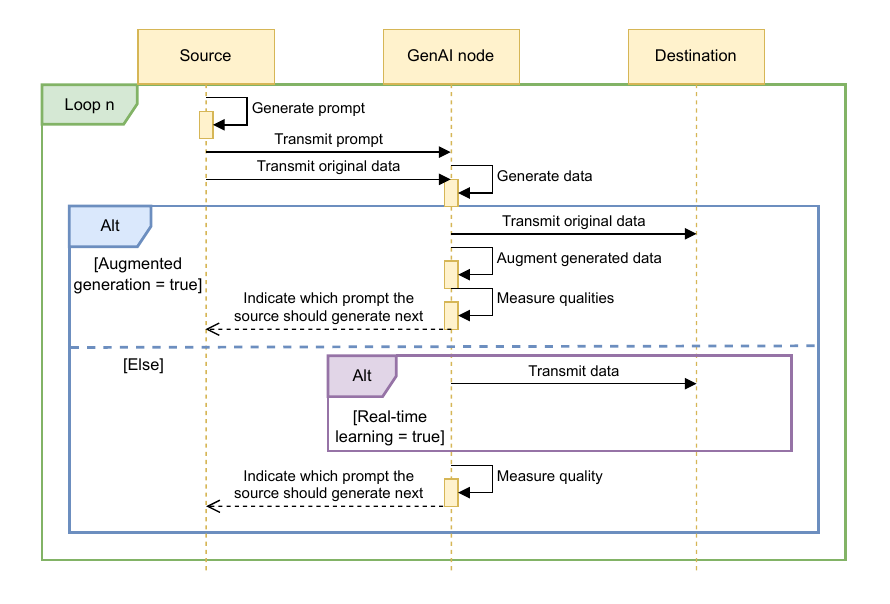}
    \caption{Message sequence diagram for the node-oriented learning process. With augmented generation, the loop represents the $N_L$ data points; The generative node $g$ creates all the necessary encoding sizes locally in each loop. Without augmented generation, the loop represents $N_p$ encoding sizes per data point.}
    \label{fig:InitProtc_node-oriented}
\end{figure}

The node-oriented learning protocol enables the GenAI node to measure quality and fit the rate-quality function, which is then returned to the source. Measuring the quality at the node requires the source to transmit the original data along with the encodings, which oftentimes  will incur an increased communication load on the source-node link. Fig.~\ref{fig:InitProtc_node-oriented} shows the message sequence of the protocol. For each data point, the source generates encodings $\pi_{(x_n,L_p)}$ and transmits them, along with the original data $x_n$, to $g$, which generates an approximation $\hat{x}_n$ for each encoding size, measures the quality $Q(\hat{x}_n)$, and fits $D(L_p,g_\theta(\cdot))$. In standard operation, the source generates $N_p$ encodings per data point. However, if the node supports augmented generation, such as the \gls{ps} method presented in \cite{thorsager2024generative}, the source will only transmit a single encoding, that is, the smallest encoding $L_\text{min} \in \mathcal{P}$ per data point. Using this one encoding, the node can generate the remaining encoding sizes by augmenting the generated data $\hat{x}_n$ with different amounts of the original data. For standard operation, the communication cost per data point is $\kappa_n=E(|x_n|) + \sum_{L_p \in \mathcal{P}_{N_p}} L_p$ for $N_p$ tested encoding sizes. Using augmented generation, the communication cost per data point is $\kappa_n = E(|x_n|) + L_{\min}$. Similar to source-oriented learning, this protocol is restricted to deviation-based quality metrics.

For node-oriented learning, an alternative method for learning the rate-quality function is possible. Since the node is receiving the original data $x_n$, it can use it to estimate the distribution of $\mathcal{X}$ and determine if it already has a pre-fitted rate-quality function that is appropriate for the source data.

\subsubsection{Destination-Oriented Learning}

\begin{figure}
    \centering
    \includegraphics[width=1\linewidth]{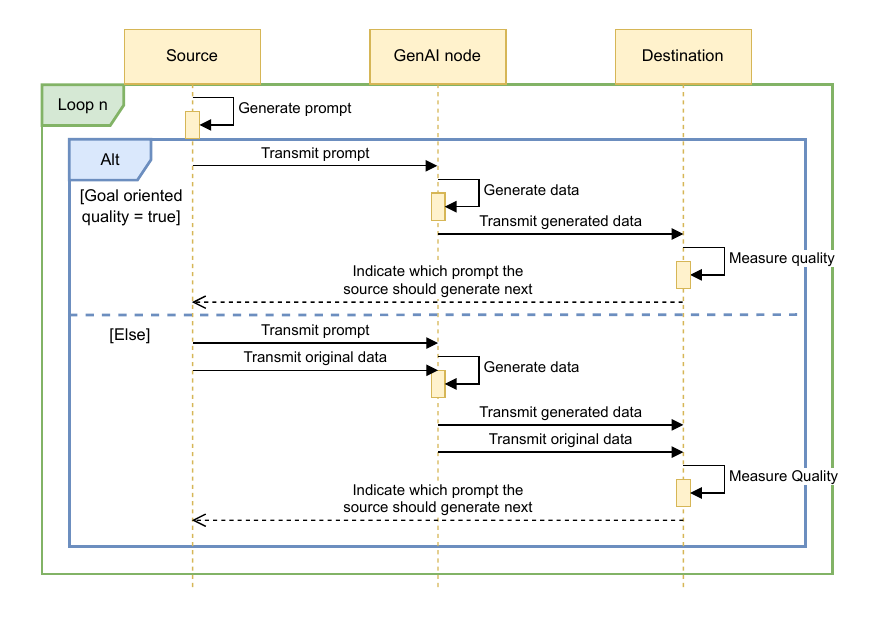}
    \caption{Message sequence diagram for the destination-oriented learning process. For each of $N_L$ data points, $N_p = |\mathcal{P}_{N_p}|$ loops are required to measure the quality for the selected encoding sizes.}
    \label{fig:InitProtc_destination-oriented}
\end{figure}

The destination-oriented learning protocol enables the destination to measure the quality and fit the rate-quality function, which is then returned to the source. This means that goal-oriented quality metrics can be used in addition to deviation-based metrics. Figure~\ref{fig:InitProtc_destination-oriented} shows the message sequence of the protocol. For each data point $x_n$, the source generates encodings $\pi_{(x_n,L_p)}$ of varying sizes $L_p \in \mathcal{P}_{N_p}$ and transmits them to $g$. The node $g$ generates an approximation $\hat{x}_n$ for each encoding size, which it transmits to $d$ for quality measurement. While destination-oriented learning would likely mainly be used with goal-oriented metrics, if deviation-based metrics are used, the source will additionally transmit the original data point $x_n$ once for all $N_p$ encodings generated from $x_n$. 

It should be noted that for some goal-oriented quality metrics, the source may not be able to choose arbitrarily many encoding sizes per data point. If the task performed at $d$ is stateful, the same data point cannot be reused with multiple encoding sizes. In this case, more unique data points are required (larger $N_L$) in order to fit the rate-quality function. 

For goal-oriented metrics, which do not require the original data point $x_n$ for the quality measurement, the communication cost per data point is the same as for source-oriented learning, i.e., $\kappa_d=\kappa_s= \sum_{L_p \in \mathcal{P}_{N_p}}(L_p + |\hat{x}_n|)$ for $N_p$ tested encoding sizes. However, there is a difference in where the communication cost occurs. For source-oriented learning, the data approximation is transmitted from $g$ to $s$, whereas it is transmitted from $g$ to $d$ in destination-oriented learning with goal-oriented metrics. For deviation-based metrics, the communication cost per data point is the same as for node-oriented learning without augmented generation, i.e., $\kappa_d=\kappa_n=E(|x_n|) + \sum_{L_p \in \mathcal{P}_{N_p}} L_p$ for $N_p$ tested encoding sizes.

\section{Initialization Procedure}\label{sec:InitProcedure}

In this section, we detail the practical considerations for the learning process assuming that the source has made a contract with a \gls{genai}-aided network node. A natural question is how to determine the number of data points to use for the estimate of the rate-quality function. Since the rate-quality function is only an estimate, it follows that the more data points the function is fitted on, the less uncertainty the source should have in its accuracy. This means that there will be a trade-off between the resources spent estimating the rate-quality function and the resource savings when the source communicates using the estimated function. Consider the case where the source is transmitting under a quality constraint. The source estimates the uncertainty in the rate-quality function based on the number of data points used in the estimate. This can be done either by calculating a bound such as the prediction interval or distribution-free tolerance limit if the rate-quality function is estimated using a traditional regression model or using tools such as conformal prediction \cite{shafer2008tutorial} if the regression model is \gls{ml}-based. 

While it may be typical to assume normally distributed residuals when calculating uncertainty bounds, for some metrics or encoding methods, the normality assumption will not hold which may result in underperformance on quality adherence. If the exact distribution of the residuals is not known, methods such as the distribution-free tolerance interval by Wilks \cite{wilks1941determination} can be used instead. This method makes no assumption on the underlying distribution of the residuals which makes it more robust in such cases. Additionally, for the tolerance interval, it is possible to compute the minimum number of required data points to meet an arbitrary value at the given probability.

The source may wish to find an optimal number of data points for the estimation by optimizing this trade-off, but this would require knowing the uncertainty in the estimates of the rate-quality function for all numbers of source data points. While one could employ statistical procedures to determine the optimal dataset size, this is outside the scope of this work. Instead, a pragmatic approach is to base the decision on a pre-allocated budget which either consists directly of a number of data points $N_L$, or can be translated into such.

\subsection{Budget-Based Initialization}\label{subsubsec:Budget}
If the budget is not directly based on the number of data points $N_L$, the source can use other constraints such as a communication or time limit and calculate $N_L$ from that.

Given a communication budget $B_C$, the source finds the maximum number of data points it can afford given the communication cost per data point $\kappa$:
\begin{equation}
    N_L = \left\lfloor \frac{B_c}{\kappa} \right\rfloor
\end{equation}
Here, $\kappa$ will take one of three values depending on the learning protocol used and is based on the costs defined in Section \ref{subsec:Learning}.

Given a time budget $B_t$, the source uses the latency model for the processing and transmission time $T_L$ of a single data point in Equation \eqref{eq:latency_model} in Section \ref{subsec:latency_model} to determine the number of data points it can process during the learning phase: 
\begin{equation}
N_L = \left\lfloor \frac{B_T}{T_L} \right\rfloor
\end{equation}
The generation time $T_G$ is provided during contracting (Section \ref{subsec:conctract}), and the communication delay $T_C$ can be estimated through probing. The communication delay depends on the amount of data transmitted, which varies by learning protocol and uses the same considerations for $\kappa$ as described for the communication budget.

If it is not possible to estimate the cost (be it time- or communication-based) prior to initiating the learning, the source can also perform a hybrid method. Here, the source uses the observed costs to evaluate if it has the available budget to process an additional data point. This is relevant in cases where the source does not have a backlog of data for the learning phase and must wait for, e.g., a new sensing event between each data point. 

Since both the time and communication budgets directly relate to the number of data points used in the estimate of the rate-quality function, in the rest of this paper, we will use the term \emph{estimation budget} (and the size of the estimation budget) to indicate this amount. When discussing the amount of data transmitted in the learning phase or the time spent, we will refer to the cost of the estimation budget or simply the cost of learning.

\subsection{Budget-Based Initialization with Pilot Transmissions} \label{subsec:pilots}
If the source is performing a single estimate of the rate-quality function, it may not be able to strictly adhere to quality constraints. While using uncertainty bounds, on average, allows a source to meet a specific quality requirement with a certain probability, this is only marginal coverage. There is no conditional coverage for any individual subset of data points used for the estimate. Furthermore, if the distribution of the source data changes, the rate-quality function will not remain accurate for the new data. To overcome this, the source can use pilot-inspired transmissions in the post-learning phase to improve the accuracy of the function beyond the initial estimate. This will increase the likelihood that the estimate accurately represents the distribution of the source data, and will allow the source to adjust the function over time in case the distribution changes. Unlike the initial learning phase, which relies on a pre-allocated estimation budget, pilots are used during the post-learning phase to continuously refine the estimate. The pilot transmissions are necessary to enable updates of the estimated rate-quality function, as the source will not get feedback from the normal post-learning data transmissions regarding the quality of the generated data. In order to enable the feedback, the pilots will consist of the same encoding transmissions as during the learning phase.

\section{Experimental Results}\label{sec:Results}

The purpose of this section is to validate the proposed learning protocol through a case study on image transmission and generation. Validating the framework requires testing it across the elements that govern its use: the data distribution at the source, the generative model, the encoding method used to construct the rate-quality function, and the quality metric applied. We address data distribution by comparing quality adherence when the source learns its own rate-quality function against when it uses a node-advertised function fit on a different data subset. We address model and encoding method by including results from two generative models HiFiC \cite{HiFiC} and MS-ILLM \cite{muckley2021neuralcompression}, and two encoding methods \gls{ps} and \gls{pe} noting where specific combinations are limited by model or metric constraints. We address quality metric by using both a perceptual metric, \gls{lpips}, and a goal-oriented metric, softPQ.

For this, we relate the rate-quality function $D(L_p,g_\theta(\cdot))$ to a perceptual quality metric used for images and to a goal-oriented metric based on inference accuracy. In particular, for the perceptual quality metric we use \gls{lpips} \cite{LPIPS}, which is a perceptual image quality measure that calculates the distance between extracted features of two images (the original and the generated image). Note that since we use a deviation-based quality metric, the quality for each encoding size in the rate-quality function is defined as $1/\delta(x_n,\hat{x}_n)$, where $\delta()$ is the \gls{lpips} function. The feature extraction is performed using the Alexnet model architecture. 

For the goal-oriented metric, we investigate an inference task by measuring softPQ using the yolo11x-seg model on the COCO2017 validation set. For each image, softPQ is obtained by matching detected segments against the ground-truth annotations and summing the Intersection over Union (IoU) of the matched pairs, normalised by the mean number of detected and annotated objects,
\begin{equation}
    \text{softPQ} = \frac{\sum_{(i,j) \in \mathcal{M}} \text{IoU}(i,j)}
                         {\tfrac{1}{2}\left(N_\text{det} + N_\text{gt}\right)}.
    \label{eq:softpq}
\end{equation}
Unlike panoptic quality, no threshold is applied to the individual overlaps, so a partially recovered object contributes in proportion to how much of it is recovered rather than being counted as a whole detection or not at all. This is beneficial for the purpose of the framework as it enables more granular evaluation of the accuracy of an inference. Since \eqref{eq:softpq} is a score rather than a distortion, and since its value on the original image differs from one image to the next, we express the quality as the softPQ relative to the original,
\begin{equation}
    Q(L_p) = \frac{\text{softPQ}(L_p)}{\text{softPQ}(|x_n|)},
    \label{eq:shortfall}
\end{equation}
which is 1 at the original by construction and grows as the encoding size is increased. As an example, $Q_\text{min}=0.75$ would be a requirement that the generated images retain at least 75\% of the detection performance as the original image.

The rate is measured as the size of the encodings used to generate the images. For the image generation, we use two separate \gls{genai}-based image compression models, HiFiC \cite{HiFiC} and MS-ILLM \cite{muckley2021neuralcompression}, using the model weights for the encoder and decoder pairs published along with their respective papers. In particular, for HiFiC, we use the weights for all 3 pre-trained encoding rates, while for the MS-ILLM model, we only use the weights for the 6 largest pre-trained encoding rates. Both models compress images into encodings in the form of latent representations. To ensure that the encoding size is invariant to the number of pixels in an image, we define it as the number of \gls{bpp} of the latents, calculated as the actual size in bits of the latent divided by the number of pixels in the original image. We use a subset of randomly selected images from the COCO2017 \cite{CoCo} image dataset, chosen in part for its limited overlap with the models' training data. MS-ILLM does not include COCO, and while HiFiC's exact training set is not stated, the authors describe using high-resolution images \cite{HiFiC}, which COCO2017's largely VGA-resolution content ($640\times480$ pixels) is unlikely to match.

\subsection{Learning Cost}\label{subsec:CaseStudy_LearningCost}

As presented in Section \ref{subsec:Learning}, the communication cost per data point is dependent on the learning protocol used. Hence, we denote the total learning cost as $K_L=N_L\kappa_i$, $i \in \{s,n,d\}$.

Considering the image data from the COCO dataset, we can quantify an example cost of each approach by considering the average image size within the chosen subset, which is 4.824 \gls{bpp}. Likewise, the size of the generated images is 4.824 \gls{bpp}. The smallest possible encoding size using the HiFiC model (when none of the original image is used for augmentation) is 0.299 \gls{bpp}. 

Theoretically, there can be an arbitrary number of encoding sizes and, therefore, there are arbitrarily many possibilities for the subsets $\mathcal{P}_{N_p}$. However, increasing the number of possibilities for $\mathcal{P}_{N_p}$ would incur a large computational cost for learning the rate-quality function. Therefore, for this case study, we consider a reduced yet representative subset $\mathcal{P}_{N_p}$ with an average encoding size $L_\text{avg} = \frac{1}{N_p}\sum_{\mathcal{P}_{N_p}}L_p=2.59$ \gls{bpp} for the results that are based on the \gls{ps} encoding method. However, due to practical considerations for the implementation of the \gls{pe} method, the average encoding size is significantly smaller at $L_\text{avg} = 0.467$ \gls{bpp}.

Given the image resolution of the dataset of $640\times480$ (VGA) (and using the encoding sizes for \gls{ps}), we get data sizes of $L_\text{min}= 0.092$, $L_\text{avg} = 0.787$, and $x_n= 1.482$ Mbits. Substituting these values into the cost formulas from Section~\ref{subsec:Learning}, we get the results shown in Table \ref{tab:learningCost}.

\begin{table}
\centering
\caption{Learning cost for the different learning options and approaches.}
\label{tab:learningCost}
\begin{tabular}{@{}l@{}lll@{}}
     \toprule
     \textbf{Learning location}& &\textbf{Cost formula} & \textbf{Specific cost [Mbits]}\\
     \midrule
    \textbf{Source} &$\kappa_s$ & $N_p(L_\text{avg} + |\hat{x}_n|)$ & $N_p2.269$\\
    \textbf{Node}\\
     \hspace{1em}Augmented  &$\kappa_n$ &  $|x_n| + L_{\min}$ & $1.574$\\
     \hspace{1em}Standard  &$\kappa_n$ &  $|x_n|+N_pL_\text{avg}$ &  $1.482 + 0.787N_p$\\
    \textbf{Destination} \\
    \hspace{1em}Goal-oriented & $\kappa_d$ & $N_p(L_\text{avg} + |\hat{x}_n|)$ & $2.269N_p$ \\
    \hspace{1em}Deviation-based & $\kappa_d$ & $|x_n| + N_p(L_\text{avg} + |\hat{x}_n|)$ &  $1.482 + 2.269N_p$\\ \bottomrule
\end{tabular}
\end{table}

\subsection{Communication Savings}\label{subsec:CaseStudy_commsSavings}

Since the source is transmitting encodings instead of the original data for the post-learning transmissions, the communication cost will be lowered by the difference between the size of the optimal encoding size and the size of the original data. We define this difference as the communication savings $W = N_Cw$, where $N_C$ denotes the number of images transmitted during the post-learning phase and $w=(|x_n|-L_P^*)$ is the per-image savings. Since the source cannot control the data source, the main aspects of this cost are the encoding strategy the source uses and how it chooses the encoding size.

The encoder-decoder models from \cite{HiFiC} and \cite{muckley2021neuralcompression} have three and six pre-trained variants with different encoder output dimensions, yielding different encoding sizes. While custom encoder dimensions enable arbitrary encoding sizes, each variant requires individual training with a matched decoder. For this case study, we employ two complementary approaches to extend the rate-quality function beyond the pre-trained model variants~\cite{thorsager2024generative}: 1) \gls{pe}, a method where we extrapolate the performance of the three pre-trained encoder-decoder variants to any arbitrary encoding size; 2) \gls{ps}, a data augmentation method where portions of the original image are transmitted alongside the encoding and composited onto the generated image at node $g$. 

The \gls{pe} method provides a theoretical means to extend the rate-quality function and, while it may not be practically viable for HiFiC or MS-ILLM in their current forms, several models have demonstrated additional extension of the rate-quality function through additional pre-trained model variants, which fit the interpolation of the rate-quality function~\cite{he2022elic,minnen2020channel}. On the other hand, \gls{ps} provides a method for extending the set of encoding sizes beyond the three pre-trained models that can be readily implemented in any image \gls{genai} model without the need for retraining. Even though there are many options for implementing \gls{ps}, for the sake of simplicity, we use only the encoder/decoder pair producing the smallest encodings from the HiFiC model for latent generation and select pixels uniformly at random. We showed in \cite{thorsager2024generative} that \gls{ps} with random selection of pixels does not perform on par with \gls{pe}. However, it may be possible through targeted selection strategies (e.g., RoI-based foreground preservation, similar to \cite{cai2019end}) to improve the performance.

While the rate-quality function gives the expected quality for a given encoding size, it does not take into account the uncertainty in the estimate of the function. To account for this, we employ Wilks' distribution-free alternative \cite{wilks1941determination}. For this method, the bound is read off the sampled prediction errors themselves rather than from a fitted density. For an encoding size $L_p$, let $e_{L_p}^{(1)} \le \dots \le e_{L_p}^{(N_L)}$ be the ordered prediction errors of the $N_L$ images used for the estimate. If the errors are drawn from a continuous distribution $F$, then the fraction of the population that the $k$-th of them covers is
\begin{equation}
    F\!\left(e_{L_p}^{(k)}\right) \sim \mathrm{Beta}(k,\, N_L-k+1),
\end{equation}
independently of $F$ itself \cite{wilks1941determination}. Requiring that this order statistic cover at least a fraction $p$ of the population with confidence at least $\gamma$ gives the condition
\begin{equation}
    \Pr\!\left[F\!\left(e_{L_p}^{(k)}\right) \ge p\right]
      = 1 - I_{p}\!\left(k,\, N_L-k+1\right) \ge \gamma,
    \label{eq:wilks}
\end{equation}
where $I_p(\cdot,\cdot)$ is the regularised incomplete beta function. Taking the smallest $k$ that satisfies \eqref{eq:wilks}, the quality of any encoding size is bounded by $\hat{Q}(L_p) + e_{L_p}^{(k)}$, which is a one-sided $p/\gamma$ tolerance bound and holds whatever the shape of the error distribution. Since \eqref{eq:wilks} can only be satisfied for $N_L \ge \lceil \ln(1-\gamma)/\ln p \rceil$, the method also sets a lower limit on the number of images the source must measure before it can select an encoding size at all. E.g. for $p = \gamma = 0.95$ this limit is $N_L = 59$ images, at which the bound is the largest observed prediction error. However, the source can freely choose the value of $\gamma$ based on its desired probability that any given realization of the rate-quality function will adhere to the quality requirement. 

For the purpose of the following tests, we aim for marginal coverage. For this, we select the order statistic from the expectation of the beta distribution, $E[F(e_{L_p}^{(k)})] = k/(N_L+1)$, which is independent of $F$, and take the smallest index reaching the target,
\begin{equation}
    k = \left\lceil p\,(N_L+1) \right\rceil .
    \label{eq:kMean}
\end{equation}
Since only $N_L$ errors are available, the index is bounded by $k \le N_L$. Substituting \eqref{eq:kMean} and solving for the budget gives
\begin{equation}
    N_{\min} = \left\lceil \frac{p}{1-p} \right\rceil ,
    \label{eq:Nmin}
\end{equation}
which for $p = 0.95$ is $N_L = 19$ images, in place of the $59$ required at $\gamma = 0.95$. The confidence $\gamma$ is the obtained by evaluating the left-hand side of \eqref{eq:wilks} at this index,
\begin{equation}
    \gamma(N_L) = 1 - I_{p}\!\left(k,\, N_L-k+1\right) ,
    \label{eq:gammaOut}
\end{equation}
reducing to  $1-p^{N_L}$ while $k = N_L$, so that the bound at the minimum budget covers $95\%$ of the population with confidence $\gamma = 0.62$.

\begin{figure}[t]
\centering
\subfloat[]{\includegraphics[width=\columnwidth]{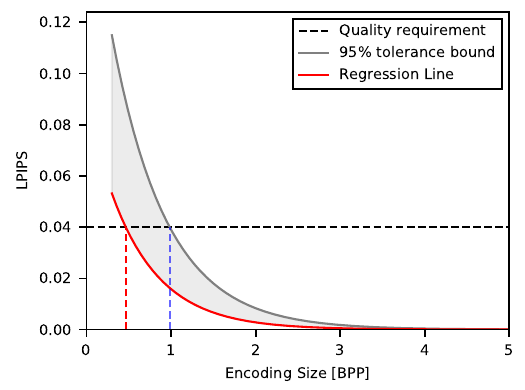}\label{fig:Wilks_PE_HiFiC_19}}\\[-1pt]
\subfloat[]{\includegraphics[width=\columnwidth]{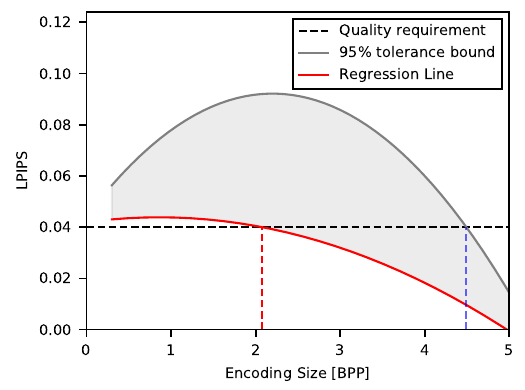}\label{fig:Wilks_PS_HiFiC_19}}
\caption{Rate-quality function for the HiFiC models using the \gls{pe} and \gls{ps} encoding methods measured with the LPIPS quality metric. The bounds are calculated using the tolerance limit with a budget of 19. The quality requirement is set to $Q_\text{min}=1/0.04=25$. Subfigures (a) and (b) show \gls{pe} and \gls{ps} respectively.}
\label{fig:Wilks_HiFiC_LPIPS}
\end{figure}

Figure \ref{fig:Wilks_HiFiC_LPIPS} shows examples of the tolerance limit with an estimation budget of 19 images for the HiFiC compression model measured with the LPIPS metric. While Figure \ref{fig:Wilks_PE_HiFiC_19} shows the expected behavior of decreasing tolerance limit width (distance from the bound to the mean) with encoding size, Figure \ref{fig:Wilks_PS_HiFiC_19} has an increased interval width at the middle encoding sizes. However, since the mean quality shows constant improvement with encoding size, the increasing interval width indicates that there are outlier images where low fractions of original pixels deteriorate the quality. While this means that the bound is non-monotonous, this does not mean that the source cannot use this rate-quality function. It will result in lower performance during the post-learning communication as the source would limit the usable region of the rate-quality function to a non-continuous set of encoding sizes. In particular, for the rate-quality function shown in Figure \ref{fig:Wilks_PS_HiFiC_19}, the usable region contains the lowest encoding size together with a disjoint range from 4.099 to 4.824 BPP\footnote{For the purpose of demonstrating the effect of non-monotonicity in the rate-quality function, we have purposefully chosen the most extreme example from the dataset. Most of the cases of non-monotonicity of the bound is less severe and the majority of the estimated rate-quality functions do not show this behavior at all.}.

\begin{figure}[t]
\centering
\subfloat[]{\includegraphics[width=\columnwidth]{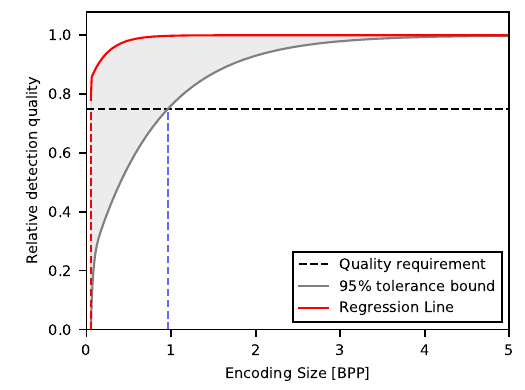}\label{fig:Wilks_PE_MS-ILLM_softPQ_19}}\\[-1pt]
\subfloat[]{\includegraphics[width=\columnwidth]{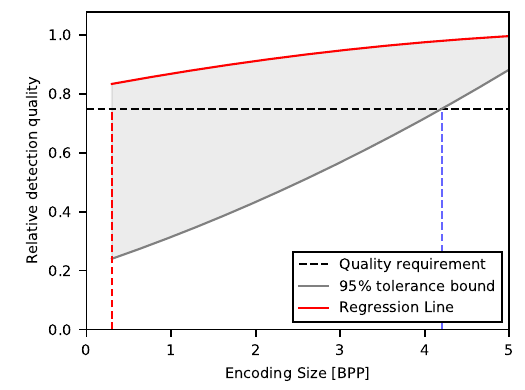}\label{fig:Wilks_PS_HiFiC_softPQ_19}}
\caption{Rate-quality function for the HiFiC model using the \gls{ps} encoding method and the MS-ILLM model using the \gls{pe} encoding method measured with the goal-oriented softPQ metric. The bounds are calculated using the tolerance limit with a budget of 19. The quality requirement is set to $Q_\text{min} = 0.75$. Subfigures (a) and (b) show MS-ILLM with \gls{pe} and HiFiC with \gls{ps} respectively.}
\label{fig:Wilks_softPQ}
\end{figure}

The Figure also shows how the source would use the rate-quality function when there is a quality constraint on the communication. Given a quality constraint of the LPIPS value 0.04 ($Q_\text{min}=25$), the rate-quality function indicates that average encoding sizes of 2.07 and 0.463 \gls{bpp} should be chosen for the \gls{ps} and \gls{pe} methods, respectively. However, to meet the quality requirement with a probability of 0.95, the source should choose an encoding size of 4.49 \gls{bpp} for \gls{ps} and 0.98 \gls{bpp} for \gls{pe}. Extending this to the rate-quality functions for the goal-oriented metric, shown in Figure \ref{fig:Wilks_softPQ}, we see a similar difference between the \gls{pe} and \gls{ps} encoding methods.

\subsection{Quality Adherence on Advertised Performance}\label{subsec:qualityAdherence_advertised}

To exemplify the impact of the proposed learning protocol, we set up a comparison against the case where a source uses an advertised rate-quality function. Per our argument on the utility of such advertised rate-quality functions, we would expect that in cases of distribution shift in the source data, the source will not meet the quality requirement when using advertised functions. We test this by splitting the COCO dataset based on the content in the images. In particular, we make three subsets: 1) Indoor images. 2) Urban images. 3) Human images. To avoid overlap between the sets, indoor and urban images with people are only added to the human images set.

\begin{figure}
    \centering
    \includegraphics[width=\linewidth]{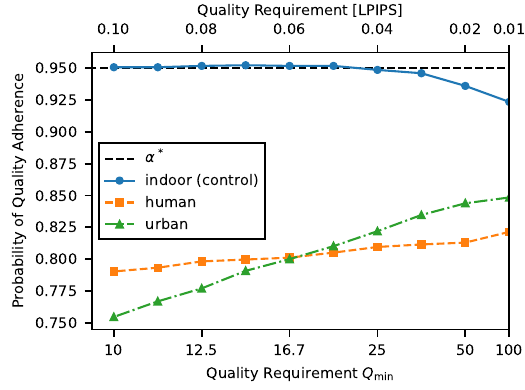}
    \caption{Quality adherence as a function of the quality requirement using an advertised rate-quality function for the \gls{pe} encoding method. }
    \label{fig:qualityAdherence_Advertised}
\end{figure}

From the quality measurements across the sets, we observe that indoor images measure higher in quality on average and with a tighter spread on the quality, confirming a measurable distribution shift between the sets. We consider a case where the generative node advertises a rate-quality function fit on the indoor subset ($\approx$ 2000 images) using the \gls{pe} encoding method with the HiFiC model. The source uses this advertised function to transmit images from the urban and human sets, and, as a control, the indoor set itself, each with $\approx 2000$ images. Figure \ref{fig:qualityAdherence_Advertised} shows the resulting quality adherence. Since the indoor images are generally generated at a higher quality, the source is more likely to choose encoding sizes that are too small, resulting in a quality adherence of $\approx 80\%$ instead of the target $95\%$ across the urban and human subsets. While the control does show slightly reduced quality adherence at the stricter quality requirement of LPIPS $< 0.03$, this is mainly due to limitations in the data used to fit the rate-quality function. At this requirement, the optimal encoding size following Equation \eqref{eq:quality_constrained_Mode} (Section \ref{subsec:communication_modes}) lies in a region of the rate-quality function far from the nearest pre-trained rate level of the HiFiC model.

This lacking quality adherence shows that the distribution shift in the source data can have a significant impact on the source's ability to communicate under quality constraints. However, while this demonstrates the importance of learning the rate-quality function for the particular data at the source, the impact of the distribution shift will not always be as significant as shown in Figure \ref{fig:qualityAdherence_Advertised}.

\subsection{Quality Adherence at Calculated Estimation Budget} \label{subsec:qualityAdherence_calculatedBudget}

This subsection validates that the minimum estimation budget of 19 images, calculated in Section \ref{subsec:CaseStudy_commsSavings}, achieves the target quality adherence $\alpha^*$ in practice. We test this across both quality metrics (LPIPS and softPQ) and both encoding methods (\gls{ps} and \gls{pe}), by measuring the empirical quality adherence over a significant number of unique realizations of the rate-quality function drawn from the COCO dataset. In particular, we make 1000 realizations of the rate-quality function estimates using unique combinations of 19 images. We then calculate the probability of adhering to a specific quality requirement for each individual estimate over a set of randomly drawn test images. By averaging the quality adherence over all realizations, we get the empirical marginal coverage probability.

\begin{figure}
    \centering
    \includegraphics[width=\linewidth]{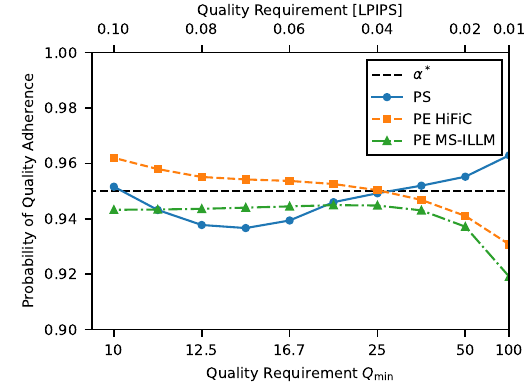}
    \caption{Quality adherence as a function of the quality requirement using a rate-quality fit on 19 images for the \gls{ps} and \gls{pe} encoding methods measured with the LPIPS metric.}
    \label{fig:qualityAdherence_LPIPS_N19}
\end{figure}

Figure \ref{fig:qualityAdherence_LPIPS_N19} shows the quality adherence on the LPIPS metric for the \gls{pe} and \gls{ps} encoding methods. It shows that the methods are mostly able to stay within 1 percentage point of the quality adherence target $\alpha^*$ in spite of the limited budget of 19 images. However, while the strictest quality requirements $Q_\text{min}>25$ (LPIPS$<0.04$) do not adhere fully and land up to 3 percentage points below the target, this is the same behavior as the adherence in Figure \ref{fig:qualityAdherence_Advertised} where the rate-quality function is fit using 2000 images. This suggests that it is not caused by the use of tolerance limit, but rather a consequence of the \gls{pe} encoding method. If the source needs stricter adherence to the quality requirement, beyond marginal coverage, it can adjust the $\gamma$ parameter of the tolerance limit to increase the probability that any given realization of the rate-quality function will show quality adherence, though this increases the budget, as shown in the example given in Section \ref{subsec:CaseStudy_commsSavings}, where $\gamma=0.95$ results in a minimum estimation budget of 59 images.

\begin{figure}
    \centering
    \includegraphics[width=\linewidth]{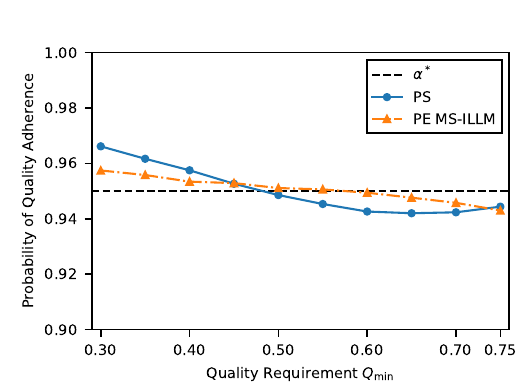}
    \caption{Quality adherence as a function of the quality requirement using a rate-quality fit on 19 images for the \gls{ps} and \gls{pe} encoding methods measured with the softPQ metric.}
    \label{fig:qualityAdherence_softPQ_N19}
\end{figure}

Figure \ref{fig:qualityAdherence_softPQ_N19} shows the corresponding quality adherence for the goal oriented softPQ metric. Unlike the LPIPS results, this figure only includes \gls{pe} with the MS-ILLM model and not the HiFiC model. When considering the goal-oriented task of inference accuracy, \gls{pe} with the HiFiC model was unable to meet the required quality adherence regardless of budget. We observed that the larger pre-trained encoding rates often performed better than the original images by lowering the probability of false positive detections, so an exponential decay rate-quality function cannot accurately describe the performance. Given the limited number of pre-trained encoding rates for the HiFiC model at 3, however, there is insufficient data to fit more complex functions, leaving results only for the MS-ILLM model using the \gls{pe} encoding method. The remaining combinations shown in Figure \ref{fig:qualityAdherence_softPQ_N19} behave consistently with the LPIPS results above, including the same $\gamma$-trade-off if stricter adherence is required.

\subsection{Performance of Framework}\label{subsec:FrameworkPerf}

The communication gains realized by the framework are not a fixed property of it, but depend directly on the encoding method and model used. To illustrate this, we calculate the communication savings $w$ with respect to lossless compression (PNG) and lossy compression (JPEG) baselines. This comparison is not intended as a precise performance benchmark of \gls{ps} and \gls{pe}, or of the HiFiC and MS-ILLM compression models, against PNG and JPEG. Rather, it illustrates whether using the framework, regardless of the specific encoding method and model, can outperform not using it, under the requirement of successfully adhering to the quality requirement. JPEG and PNG are lightweight, fast compression methods that can run on UE devices without requiring the framework, which is why they serve as the non-framework baselines here.

We define the viability point $N_V$ as the minimum number of post-learning transmissions where the learning cost is recovered: $K_L \leq N_Vw$. For JPEG, we estimate its rate-quality function using 1000 images with the same tolerance limit approach as for \gls{ps} and \gls{pe}. However, since the source can measure the quality without any transmissions to and from $g$, there is no communication cost during learning.

Table \ref{tab:wilksLpips} compares communication costs against the PNG and JPEG baselines. Against PNG, the results confirm that the rate-quality functions estimated using the proposed framework enable the source to communicate more efficiently. Regardless of the quality requirement $Q_\text{min}$ the source is able to choose an encoding size that outperforms PNG which allows it to see communication gains after 7-9 images for the \gls{pe} encoding method and 53-225 images for the \gls{ps} method. This difference in performance between the encoding methods is more apparent in comparison with the JPEG baseline where the \gls{ps} encoding method fails to realize communication gains across all quality requirements. This is in contrast to the \gls{pe} encoding method which for both HiFiC and MS-ILLM outperforms JPEG and realizes gains consistently after around 20 images regardless of quality requirement. This reveals an important practical consideration: communication savings depend on the compression method chosen, not just the protocol's ability to characterize it.

\begin{table*}[t]
    \centering
    \caption{Learning cost, data savings and viability point under the distribution-free tolerance limit, for the perceptual metric.}
    \label{tab:wilksLpips}
    \begin{tabular}{clcccccc}
    \hline
        Method                                & $Q_\text{min}$  & Budget   & $K_L$      & $w$ (PNG)   & $N_V$ (PNG)   & $w$ (JPEG)   & $N_V$ (JPEG) \\ \hline
        \multirow{4}{*}{\gls{pe} (HiFiC)}    & $33.33$         & 19      & 26.62      & 3.467       & 8             & 1.497        & 18 \\
                                              & $25$            & 19      & 26.62      & 3.672       & 7             & 1.400        & 19 \\
                                              & $20$            & 19      & 26.62      & 3.832       & 7             & 1.329        & 20 \\
                                              & $14.29$         & 19      & 26.62      & 4.072       & 7             & 1.230        & 22 \\ \hline
        \multirow{4}{*}{\gls{pe} (MS-ILLM)}  & $33.33$         & 19      & 35.73      & 3.927       & 9             & 1.964        & 18 \\
                                              & $25$            & 19      & 35.73      & 4.083       & 9             & 1.822        & 20 \\
                                              & $20$            & 19      & 35.73      & 4.203       & 9             & 1.714        & 21 \\
                                              & $14.29$         & 19      & 35.73      & 4.381       & 8             & 1.553        & 23 \\ \hline
        \multirow{4}{*}{\gls{ps} (HiFiC)}    & $33.33$         & 19      & 97.33      & 0.432       & 225           & -1.539       & N/A \\
                                              & $25$            & 19      & 97.33      & 0.683       & 143           & -1.590       & N/A \\
                                              & $20$            & 19      & 97.33      & 0.979       & 99            & -1.523       & N/A \\
                                              & $14.29$         & 19      & 97.33      & 1.840       & 53            & -1.002       & N/A \\ \hline
    \end{tabular}
\end{table*}

\begin{table*}[t]
    \centering
    \caption{Learning cost, data savings and viability point under the distribution-free tolerance limit, for the goal-oriented metric.}
    \label{tab:wilksSoftpq}
    \begin{tabular}{clcccccc}
    \hline
        Method                              & $Q_\text{min}$  & Budget   & $K_L$      & $w$ (PNG)   & $N_V$ (PNG)   & $w$ (JPEG)   & $N_V$ (JPEG) \\ \hline
        \multirow{5}{*}{PE (MS-ILLM)}       & $0.75$          & 19      & 127.39     & 3.668       & 35            & 1.265        & 101 \\
                                            & $0.65$          & 19      & 127.39     & 4.064       & 31            & 1.179        & 108 \\
                                            & $0.55$          & 19      & 127.39     & 4.318       & 30            & 1.070        & 119 \\
                                            & $0.45$          & 19      & 127.39     & 4.488       & 28            & 0.931        & 137 \\
                                            & $0.35$          & 19      & 127.39     & 4.598       & 28            & 0.799        & 159 \\ \hline
        \multirow{5}{*}{PS (HiFiC)}         & $0.75$          & 19      & 1075.92    & 0.789       & 1364          & -1.614       & N/A \\
                                            & $0.65$          & 19      & 1075.92    & 1.529       & 703           & -1.356       & N/A \\
                                            & $0.55$          & 19      & 1075.92    & 2.267       & 475           & -0.981       & N/A \\
                                            & $0.45$          & 19      & 1075.92    & 2.970       & 362           & -0.587       & N/A \\
                                            & $0.35$          & 19      & 1075.92    & 3.437       & 313           & -0.362       & N/A \\ \hline
    \end{tabular}
\end{table*}

Table \ref{tab:wilksSoftpq} shows that the framework yields similar savings when applied to goal-oriented tasks. While the learning cost is higher, causing a shifted viability point, the source is nonetheless able to use the proposed framework to estimate the rate-quality function and apply it to realize communication gains. It should be noted that the significant increase in learning cost for the \gls{ps} encoding method when using with the goal-oriented metric stems from the cost of transmitting an increased number of encodings to the destination, most of which are larger in size as compared to the encodings from the \gls{pe} encoding method.

The case study demonstrates the protocol using two methods with distinct characteristics. \gls{ps} represents an approach that can be implemented immediately without assumptions about training multiple encoder variants, providing robust empirical validation that the protocol successfully enables estimation of rate-quality functions and enables quality-constrained transmission. \gls{pe} demonstrates scenarios where \gls{genai} compression can outperform traditional baselines, showing the protocol's utility when paired with higher-performing methods. As examples of encoding methods that adhere to the principles outlined in Section \ref{subsec:preconditions}, these results validate the protocol's method-agnostic design. The protocol provides the estimation framework regardless of compression performance, while communication savings outcomes depend on the specific models and strategies employed.

This property extends beyond the image compression case study presented here. As argued in Section \ref{subsec:preconditions}, the protocol's use of statistical sampling to estimate rate-quality functions carries no inherent dependence on data modality, provided the encoding method satisfies the stated correlation requirement. Combined with the empirical results across two compression models, two encoding methods, and two quality metrics, this supports the protocol's broader applicability beyond the specific choices made in this case study. The performance of the framework therefore comes down to the specific cost of learning the rate-quality function, the possible savings given the encoding and decoding performance, and the number of data points the source transmits after learning. As \gls{genai} compression methods continue to evolve, the protocol provides a consistent framework for sources to characterize and utilize any model and encoding strategy pair.

\section{Conclusion} \label{sec:conclusion}

This paper presented an initialization protocol for learning a rate-quality function that enables the use of intermediate \gls{genai} nodes for compression in communication networks. Unlike traditional compression, where sources can locally evaluate quality, \gls{genai}-based compression requires sources to rely on network nodes equipped with generative models both for learning the compression characteristics and for facilitating efficient communication. This necessitates the design of an initialization protocol that accounts for the communication and computational costs of the learning process itself.

The protocol defines three learning variants: source-oriented, node-oriented, and destination-oriented. These variants establish different messaging requirements and place different communication and computational loads on different network elements based on where quality measurements are performed. To illustrate the protocol's operation, we presented one possible learning approach based on a calculated minimum estimation budget using the distribution-free tolerance limit. Our experimental validation demonstrated that the framework successfully enables rate-quality function estimation and quality constrained operation across different \gls{genai}-based compression models, encoding methods, and quality metrics, supporting the framework's broader applicability beyond this specific case study. Combined with the formal argument in Section~\ref{subsec:preconditions} that the framework is agnostic to data modality provided the encoder and decoder are jointly trained and the encoding method is designed such that quality is correlated with encoding size, these results support the framework's design goal of being agnostic to the specific data, method, and model used. The results showed that the source could estimate rate-quality functions using the calculated minimum estimation budget while adhering to marginal coverage on arbitrary quality requirements, and subsequently realize communication gains over JPEG after around 20 images post-learning for the perceptual quality metric and more than 100 images for the goal-oriented metric. In cases where the source expects distribution shifts of the source data, it can make use of pilot transmissions to continuously update its estimate of the rate-quality function. However, specifying the exact pilot transmission behavior is left for future work. 

The initialization protocol provides the foundation for deploying \gls{genai}-based compression in practical communication systems, enabling sources to make informed decisions about encoding sizes while accounting for both quality requirements and the statistical uncertainty inherent in rate-quality function estimation.

\bibliographystyle{IEEEtran}
\bibliography{sources}

\end{document}